\documentclass{article}
\usepackage{jmlr2e}
\usepackage[utf8]{inputenc}

\usepackage{microtype}
\usepackage{graphicx}
\usepackage{subfigure}
\usepackage{booktabs} 

\usepackage{hyperref}

\hypersetup{
    colorlinks = true
    }

\usepackage{xcolor}

\ShortHeadings{Ideas for Improving the Field of Machine Learning}{Sodhani et al.}
\firstpageno{1}

\begin{document}

\title{Ideas for Improving the Field of Machine Learning: Summarizing Discussion from the NeurIPS 2019 Retrospectives Workshop}

\author{Shagun Sodhani,\textsuperscript{1} \hspace{.5mm} Mayoore S. Jaiswal,\textsuperscript{2} \hspace{.5mm} Lauren Baker,\textsuperscript{3} \hspace{.5mm} Koustuv Sinha, \textsuperscript{1,4,5} \hspace{.5mm} Carl Shneider,\textsuperscript{6} \hspace{.5mm} Peter Henderson,\textsuperscript{7} \hspace{.75mm} Joel Lehman,\textsuperscript{8} \hspace{.75mm} Ryan Lowe \textsuperscript{5,9} \thanks{All authors contributed equally. Contact at: ml.retrospectives@gmail.com}
\\
\\
\textnormal{
\textsuperscript{1} Facebook AI Research \\
\textsuperscript{2} IBM \\
\textsuperscript{3} Applied Invention \\
\textsuperscript{4} McGill University \\
\textsuperscript{5} MILA \\
\textsuperscript{6} Dutch National Center for Mathematics and Computer Science (CWI) \\
\textsuperscript{7} Stanford University \\
\textsuperscript{8} Work done while at Uber AI \\
\textsuperscript{9} OpenAI \\
}
}
\date{July 13, 2020}

\maketitle

\begin{abstract}
This report documents ideas for improving the field of machine learning, which arose from discussions at the ML Retrospectives workshop at NeurIPS 2019. The goal of the report is to disseminate these ideas more broadly, and in turn encourage continuing discussion about how the field could improve along these axes. We focus on topics that were most discussed at the workshop: incentives for encouraging alternate forms of scholarship, re-structuring the review process, participation from academia and industry, and how we might better train computer scientists as scientists. Videos from the workshop can be accessed at~\cite{mlretro2020}.

\end{abstract}

\section{Introduction}

One of the main goals of scientific research is to contribute to our understanding of the world. The primary way that scientific knowledge gets disseminated in most fields is through academic papers, written by researchers. However, most researchers --- whether professors, students, or employees of industry labs --- get credit for the number of papers that are accepted to prestigious conferences or journals. For professors, the number of accepted papers determine whether they will get tenure; for students, it determines their graduation times and job prospects. 

Thus, the incentives of individual researchers (writing papers that get accepted to prestigious conferences, among other things) is not the same as the behavior we want from the scientific field as a whole (producing and sharing interesting and useful scientific knowledge). This mismatch leads to a variety of problems in scientific communities, from replication crises in fields like psychology \citep{john2012measuring}, to poor scholarship in machine learning (ML) papers \citep{lipton2018troubling}. This mismatch has been exacerbated in machine learning by the rapid growth of the field, as indicated by the explosion of arXiv papers and ML conference attendees, leading to the perception of increased competition. 

The goal of the NeurIPS 2019 Retrospectives workshop\footnote{\href{https://ml-retrospectives.github.io/neurips2019/}{https://ml-retrospectives.github.io/neurips2019/}}~(workshop videos can be viewed at~\cite{mlretro2020}) was to experiment with a new kind of scholarship: retrospectives. A retrospective is a document, like a blog post, that describes authors' thoughts about their past paper that were not present in the original work.\footnote{For a more detailed description of the motivation for retrospectives, see: \href{https://thegradient.pub/introducing-retrospectives/}{https://thegradient.pub/introducing-retrospectives/}.} The workshop also accepted meta-analyses --- papers which reflect on the state of a sub-field as a whole, including disseminating newly-emerging consensus or conflict or sharing practical advice for training models or tuning hyperparameters. This led to a fruitful workshop with many interesting talks and discussions.

Central to the workshop was a panel and a brainstorming session. The topics discussed during these sessions were not confined to retrospectives and meta-analyses; rather, they spanned a variety of concerns with the current state of the field of machine learning. This report summarizes many of the ideas that were discussed during these sessions, with the goal of disseminating them more broadly and encouraging further discussion of these issues in the machine learning community. While we primarily present ideas discussed in the workshop, in some cases we point out limitations to the implementation of these ideas. However, the intent of this report is not to provide a thorough analysis of these ideas, nor to provide specific recommendations, nor to examine how these ideas are reflected in other scientific fields outside of ML. Many of the ideas in this report were presented by individuals who are not authors of this report, but were instead participants in the panel or brainstorming sessions; we acknowledge these individuals at the end of the report. 

This report is by no means an exhaustive list of the ways the field of machine learning could be improved. Important topics not discussed in this report include: increasing the inclusivity and diversity of researchers the field, paper reproducibility, training ML researchers in ethics to better understand the potential harms of ML technologies, increasing collaboration between ML researchers and other scientific disciplines, and more. 

The report is organized as follows. In Section \ref{sec:retrospectives}, we discuss the lack of incentives for alternate forms of scholarhsip, such as retrospectives and meta-analyses. In Section \ref{sec:review}, we discuss ideas for re-structuring the review process. In Section \ref{sec:industry}, we discuss the interaction between academia and industry in machine learning research. In Section \ref{sec:science}, we discuss ways to better train computer scientists in the scientific process. 
Finally, we conclude in Section \ref{sec:conclusion}.



\section{Incentivizing openness and alternate forms of scholarship}
\label{sec:retrospectives}


\paragraph{The problem.} 
One of most visible symptoms of the aforementioned misalignment between the goal of the ML field and the incentives of individual researchers is paper obfuscation. It has been noted elsewhere~\citep{lipton2018troubling} that researchers often add equations to a paper that are not necessary for understanding the content of the paper. This is a specific symptom of a broader set of behaviors, including omitting empirical results on certain datasets if they are not convincing, cherry-picking qualitative results, and setting aside hyperparameter tuning for baseline models~\citep{rendle2019difficulty}.

One idea for encouraging openness about a paper's limitations is by publishing a retrospective. A retrospective of a paper is a commentary, written by one of the authors of the original paper, that reflects on the work with the benefit of hindsight. It may include some lessons learned after the paper was published, new insights inspired by follow-up work, or details that were missed in the original work due to paper length. Authors could also use retrospectives to report intuitive ideas that did not work in practice. This information is very valuable for follow-up works, but currently, there isn't a framework to systematically disseminate such information. Hence, these insights remain within the author's network, mostly disseminated by personal conversations. Authors may be more willing to write retrospectives as they can be written \textit{after} the acceptance of the original paper, whereas authors might fear including this information in the original paper would be grounds for rejection.

In a similar spirit of self-reflection, the retrospectives workshop also accepted meta-analyses. Meta-analyses are different than review papers: a review paper aims to summarize and synthesize a wide range of papers in a specific subfield, with the goal of being very thorough, as well as providing some insights as to how the papers relate to each other. On the other hand, the goal of a meta-analysis paper isn’t to summarize the content of papers, but rather to discuss and analyze an interesting aspect of a set of papers (e.g. evaluation methodology, conflicting claims, etc.), or give an opinion about emerging trends. A meta-analysis paper doesn’t have to be thorough (it could discuss only a few papers) or limited to a narrow subfield (it could analyze broader trends across the ML community). In the submission process, meta-analyses will be unlike most other papers, with a lack of novel results. 


Retrospectives and meta-analyses, though, take time to write. Some researchers may decide to write these kinds of works in a personal blog post. However, there is a lack of formal venues that accept these kinds of works, which limits the incentive for a researcher to spend the time writing them. The workshop discussion was centered around how we can provide additional incentives for researchers to produce this kind of scholarship, or to simply be more open about the limitations of their papers. 


\paragraph{The discussion: Encouraging retrospectives and meta-analyses. }

 
One suggestion at the workshop highlighted the need for numerous smaller venues, akin to the size of some NeurIPS workshops. These smaller, closer-knit communities call for more conscientious members and might catch errors in papers that might go unseen in a larger group setting. It might be easier to discuss a retrospective in smaller communities.  Having a recurring workshop across conferences that documents its proceedings could be helpful but could create an unhealthy two-tier system. The community could force all papers to make their results open or write retrospectives. However, in that scenario, authors may do the bare minimum to satisfy the requirement, and it could lead to resentment and be ultimately counter-productive.


Several other ideas were floated: conferences request authors of established papers (e.g.\@ those that have won a best paper award\footnote{Note though that we do not endorse the existence of conference `best paper awards' in general, as a mechanism for incentivizing high-quality papers, as the selection criteria are often arbitrary.}) to write such commentary; as their work has already gained recognition, they may be more willing to be open. Seniors in the community could encourage transparency and honest writing in papers, enforce those rules in their organizations and emphasize the long-term benefits. Conference organizers could randomly sample accepted papers and ask them to write a retrospective. If this is not implemented correctly, there could be a misuse of such power. Conferences could ask for insights and practicality of proposed methods in the submission form. To integrate the idea of retrospectives into the original paper, the authors could use the discussion section more than how it's currently used in machine learning papers. To support this, conferences could provide recognition for the ``best discussion of limitations". Another idea was to either eliminate best paper awards or have lots of them for different categories. With hundreds of papers being published in a conference, the idea of ``best paper" may be misleading. Rather celebrating different aspects of a paper, such as ``best discussion of limitations", ``best comparison of results", ``best literature review", etc. may make the community more inclusive and open.

One concern raised during the workshop was that publishing mistakes in a retrospective may hurt the career of more junior members of our community. This situation can be normalized if the more senior researchers start to write retrospectives of their work, thus encouraging junior members of our community to write retrospectives.

The suggestions for encouraging meta-analysis papers centered around changes that conferences could implement, such as adding a separate track for submissions. A new track would show support for researchers spending time analyzing trends in past papers, instead of focusing on new experiments. The lack of novel results in a meta-analysis raises the possibility of these submissions being overlooked without a different set of criteria for evaluation.

In the brainstorming discussion on meta-analyses, one participant mentioned a paper track at the IEEE Symposium on Security and Privacy, called Systematization of Knowledge (SoK). Similar to the meta-analysis track at the NeurIPS 2019 workshop, this call for papers seeks to generate discussion on existing works and subfields, and submissions do not need to contain novel research results. The SoK project has more in common with survey papers than the meta-analysis track at the workshop, because part of the objective is to summarize existing research. However, unlike some criteria for survey papers, submissions must bring a new perspective by critically examining an unspoken rule, proposing a new taxonomy, or evaluating competing approaches to a problem\footnote{https://www.ieee-security.org/TC/SP2020/cfpapers.html}.

Fellow discussants agreed that the machine learning field would benefit from a project analogous to this one. While describing the motivation for the project, a Frequently Asked Questions page for SoK\footnote{http://oakland31.cs.virginia.edu/cfp.html} notes,``our community seems to lose memory of things that have been done in the past and produces too many incremental results that don't always lead to better general understanding," and workshop participants noted the same problems in ML research. Returning to the suggestion of a track for meta-analysis submissions at ML conferences, when the discussion expanded to include the SoK project (which only accepts a small number of papers each year), it was suggested that accepted papers be given an automatic oral presentation. Additionally, this track could have a longer paper length to be more conducive to extended discussions, and offer a second round of review to promising papers (a practice adopted by the IEEE symposium). Participants felt these strategies would encourage researchers to contribute to the discussion on the current state of the ML field.

The main consensus on this topic was that there is a lack of alternate styles of papers, such as retrospectives and meta-analyses. By encouraging these various formats of knowledge dissemination, we can cultivate a diversity of thought and nurture an ecosystem where various kinds of academic papers thrive as all these papers help us to push the field of machine learning forward.

\paragraph{The discussion: Other ways to increase openness.}

A separate brainstorming session focused on ways to increase transparency other than retrospectives. It was noted that, rather than writing a retrospective, researchers could simply update the original version of the paper. While this is often done to add new results, it is rarely done for sharing the authors' updated views about their paper. 
To encourage a culture of more ongoing updates to a research paper, the panel came up with an idea of citable paper versions. arXiv is one of the key platforms of dissemination of research. While it supports paper versioning, the versions themselves are not individually citable. The panel suggested that it would be useful to have citable versions which act like Git commits, which can be referenced back to other papers. This could also incentivize retrospectives, as authors writing a retrospective could add content to their original paper which could be individually cited. 
However, it remains to be seen whether crediting citations to an updated version of the paper would lead to more net citations for that paper than simply updating the paper in the current paradigm. 
More broadly, the idea of an open-source, `Git-style version tracking' approach to machine learning research is appealing in that it could promote a more collaborative approach to conducting research. Similar ideas, such as the AI Open Network\footnote{https://ai-on.org/}, have already been tried to mixed success. 

Several other ideas were discussed for increasing openness and transparency in ML. An easy-to-adopt suggestion was to encourage paper authors to have a blog post accompanying their paper. A blog post is very informative because it is the sole point of view of the first author, who is not necessarily the only author shaping the original paper. Thus, a blog post can contain critical implementation notes which are not added to the original paper, mostly due to lack of space and time. Paper blog posts are becoming increasingly popular, as they increase the exposure of a paper, 

Another very relevant aspect of transparency in research is the reproducibility of papers. This has gained more attention in the last couple of years, as there has been a strong movement to improve reproducibility in machine learning papers with the promotion of ML reproducibility checklist\footnote{\href{https://www.cs.mcgill.ca/~jpineau/ReproducibilityChecklist.pdf}{https://www.cs.mcgill.ca/~jpineau/ReproducibilityChecklist.pdf}} and Reproducibility Challenge \citep{ICLR2018Reproducibility2018, Pineau:2019, Sinha:2020,pineau2020improving}, or even reproducible code environments (CodeOcean,\footnote{\href{https://codeocean.com}{https://codeocean.com/}} CodaLab\footnote{\href{https://codalab.org/}{https://codalab.org/}}) which can be packaged together to ensure exact reproducibility. Not long after the panel discussion, PapersWithCode\footnote{\href{https://paperswithcode.com/}{https://paperswithcode.com/}} released the ML Code Completeness Checklist,\footnote{\href{https://github.com/paperswithcode/releasing-research-code}{https://github.com/paperswithcode/releasing-research-code}} which encourages conference submissions to adhere to a detailed repository readme file. Both code environment packages and the checklist serve the same purpose of ensuring enhanced transparency in ML experimentation.

Reproducibility shares many similarities with retrospectives in that both reflect on a paper with a focus of doing good science, evaluate if the hypothesis was reasonable, validate the hypothesis, assess the process of the proposed methodology without hiding any flaws, use better statistics to understand results, and evaluate the paper in hindsight. The key difference is that a paper is generally reproduced by a third-party. Due to the amount of effort and time needed by the third-party to re-implement a paper and reproduce the results, such efforts need to be recognized and given the correct incentives. Recently, formal publications that accept replications of previous research have been created, notably ReScience;\footnote{https://rescience.github.io/} however, this is not a widespread form of scholarship, and we should seek to find ways to enhance and amplify these efforts.

\section{Re-structuring the review process}
\label{sec:review}

\paragraph{The problem.}
Most scientific communities use some form of a review process to evaluate papers on their scientific merits (like correctness, applicability, trade-offs, etc.). Passing the process adds credibility to work. 
In the ideal situation, the submitted work would be reviewed by a team of quality reviewers who would provide useful feedback for improving the work and decide if the work is ready for sharing more broadly. The authors would use the reviews to improve their work, and science would progress. 

However, the review process has several flaws, and 
``reviewing the review process" was frequently brought up during the workshop. 
With the rapid growth of the machine learning field, the rate of submission of papers has far outpaced the rate at which quality reviewers are added to the pool. As a result, the quality (and usefulness) of reviews has decreased.

The fast-paced ML research culture is bringing out the inherent limitations of the review process. For instance, the culture of ``crushing the benchmark” (requiring that papers be both ``novel'' and ``state-of-the-art'' to be accepted) leads to many of the issues in paper scholarship pointed out in \citep{lipton2018troubling}. These criteria are often not correlated with scientific quality of the paper; anecdotally, many of the most influential papers contain some form of negative results.\footnote{This was stated, for example, by Prof. Yoshua Bengio during the panel.}

The over-reliance on ``number of published papers” as the metric to measure the merit of a researcher’s work is also affecting the mental well-being of students and practitioners. The sense of perpetual competition can make the research culture fearful and toxic. It prompts practitioners to churn out as many papers as possible; these submitted papers are often incomplete, thus straining the already overloaded review system. The famous NeurIPS 2014 experiment\footnote{See: http://blog.mrtz.org/2014/12/15/the-nips-experiment.html.} highlighted that ``most papers at N[eur]IPS would be rejected if one reran the conference review process (with a $95\%$ confidence interval of 40-75$\%$)”. The review process has shown no signs of becoming less noisy, thus authors have an incentive to submit the rejected work for the next cycle of review, hoping to find more amenable reviewers which further strains the system. 

\paragraph{The discussion.}
The panel discussion noted the over-reliance on the conferences (to measure the worth of research work) and the need to rethink the publication structure. Prof. Yoshua Bengio proposed an alternate arrangement where research is put up on arXiv and submitted to journals. The paper is available publicly while it is being reviewed at a journal without worrying about the cycle of deadlines. The conferences can then pick papers from the journals, and the role of conferences is to broadcast good papers rather than to filter good papers. This arrangement can reduce the anxiety that comes with the conference deadlines. An example of this is TACL, where the paper accepted in TACL can be presented at ACL, EMNLP, or NAACL. Another option is rolling deadlines. For example, UbiComp invites papers published by the ACM journal on Interactive, Mobile, Wearable and Ubiquitous Technologies (IMWUT) which has $4$ deadlines in a calendar year\footnote{http://ubicomp.org/ubicomp2020/cfp/papers.html}. The multiple deadlines alleviate unnecessary stress on the researchers. 


There was a concern that practitioners are starting to review much sooner in their careers and can easily make mistakes (even when acting in good faith) due to the lack of training. Lack of experience makes the review process very noisy, thus reducing the value of the review. During the panel, Jonathan Frankle added that in some computer science fields, senior members spend a lot more time reviewing the papers.  Several ideas were proposed to counter this problem. First, workshops and training sessions can be organized to coach new reviewers. For example, CVPR 2018 had a workshop called ``Good Citizens of CVPR”~\citep{cvprworkshop2018}, which focused on how to be ``A good CVPR reviewer[,] A good Area Chair[, and] A good author”. Similarly, CVPR 2020 had a tutorial on ``How to write a good review"~\citep{cvprworkshop2020}. Second, professors and senior researchers could train their students to be competent reviewers by helping them write mock reviews for papers or by covering relevant training material as part of the coursework~\citep{sambowmancourse}.

Another suggestion was to acknowledge good reviewers by providing Best Reviewer Awards (as already done by NeurIPS). Making reviews public for all the papers could help to fix lack of accountability on the part of the reviewer and is already done at some conferences (e.g., ICLR). It was also suggested that the reviewers' identity should be made public after the passage of time, though this could also lead to a potential rivalry between authors and reviewers.

Other suggested measures include: \textbf{i)} Capping the number of submissions by any author, to reduce the workload for the reviewers (AAAI experimented with this strategy\footnote{See, e.g., AAAI-21 call for papers where ``Each individual author is limited to no more than a combined limit of 15 submissions to the AAAI-21 technical track.'' \href{https://aaai.org/Conferences/AAAI-21/aaai21call/}{https://aaai.org/Conferences/AAAI-21/aaai21call/}}); \textbf{ii)} Increasing the pool of reviewers by requiring all authors of the submitted papers to help as reviewers if required (EMNLP experimented with this idea\footnote{See, e.g., ``Although recent global affairs will have an effect on submissions, we still anticipate another record year in terms of numbers of submissions, and thus need an ever larger troupe of qualified, engaged and committed individuals to serve as reviewers on the PC to handle these papers... For this reason, in EMNLP 2020 we are introducing a new policy: In order to submit paper(s) to EMNLP, you must nominate at least one author to serve as a reviewer, (usually the most senior author) and for that author to take on a full load (of up to 6 reviews).'' \href{https://2020.emnlp.org/blog/2020-04-26-reviewing-policy}{https://2020.emnlp.org/blog/2020-04-26-reviewing-policy}}); \textbf{iii)} Using a single reviewer (someone senior) to screen the submissions and flag out the extremely weak submissions. This way, reviewers have more time to focus on the remaining submissions. NeurIPS 2020 has begun experimenting with some of these techniques\footnote{https://medium.com/@NeurIPSConf/getting-started-with-neurips-2020-e350f9b39c28}, which have received mixed feedback, particularly on desk rejections of weak submissions.\footnote{https://syncedreview.com/2020/07/16/poorly-explained-neurips-2020-desk-rejects-peeve-ml-researchers/}

\section{Participation from academia and industry}
\label{sec:industry}

\paragraph{The problem.}

Academic institutions worldwide are struggling to retain their top scientists.\footnote{See, e.g., ``
`We can't compete': why universities are losing their best AI scientists'' at
\href{https://www.theguardian.com/science/2017/nov/01/cant-compete-universities-losing-best-ai-scientists}{https://www.theguardian.com/science/2017/nov/01/cant-compete-universities-losing-best-ai-scientists}} This portends negative ramifications not only for academic research but also for the education of future AI talent. Although the quality of basic AI research in academia is now on par with industrial AI labs, a significant portion of fundamental AI research continues to be performed in industry. The main reasons for the relatively low retention rate of AI researchers in academia include: 1) gross disparity of salary in academia and industry, 2) disproportionate amount of time spent on grant writing in combination with a high teaching load inhibits actual research time and inordinately shifts the scientist’s role from that of ‘explorer/discoverer’ to ‘manager’, and 3) academic institutions do not make it easy for academicians to dually serve as advisers or consultants to companies. 

\paragraph{The discussion.}
A view shared among some workshop participants was that the best way forward is via governments working with industry through public/private partnerships to promote an economically sustainable, virtuous AI cycle, while ensuring a free basic research base that is independent of industry interests. There are many advantages for companies to invest in institutional AI infrastructure with ‘no-strings-attached,’ besides maintaining a good public relations policy. Companies already invest a lot of their time, money, and resources to vet, screen, and train scientists. Moreover, companies compete on a rolling basis for first access to top talent that has not yet entered the job market. Academic institutions continuously supply a quality AI cohort that can be approached by companies in the form of institutionally supported career fairs for those companies which have fairly contributed to a given institution as determined by that institution. On a more individual scale, some companies already provide opportunities for thesis internships and research experience at most academic levels. However, these are relatively few and specialized. Companies also will profit from the democratization of AI – investing in girls’ and women’s programs for coding and investing in AI development in poorer countries as described, for instance, in the MIT Technology Review 2019 article entitled “The future of AI research is in Africa”.\footnote{\href{https://www.technologyreview.com/2019/06/21/134820/ai-africa-machine-learning-ibm-google/}{https://www.technologyreview.com/2019/06/21/134820/ai-africa-machine-learning-ibm-google/}}

To enable the valorization of academic research,~\citet{mendrik2019framework} advocate for a Kaggle\footnote{\href{https://www.kaggle.com/}{https://www.kaggle.com/}}-like direction: 

\begin{quote}
    [D]eployment challenges are perhaps particularly well suited to be set-up by companies, as part of their clinical trials or cohort studies. This could aid in bridging the gap between industry and academia. If data, truth criteria, and metrics are representative of the problem, direct assessment of algorithms from academia could result in more practical and accelerated use of academic algorithms.
\end{quote} 

To facilitate this process, there needs to be an infrastructure that supports academics with funding for both benchmark and algorithm submission and their dissemination. A system\footnote{An example of such a system is Eyra Nova, the non-profit part of the Enhance-Your-Research Alliance (EYRA) company.} is envisioned where benchmarks would be submitted to a platform in direct analogy to submitting a paper to a journal, and receiving credit for it would mirror a paper citation index. The submitted benchmark would subsequently be reviewed via a rating scheme which would contain suggestions for improvement and would provide a better understanding of the quality of the benchmark itself. The same would apply for algorithm submission: the algorithm would be submitted to such benchmarks on the platform, just like submitting a paper to a journal and obtaining citation value for it. Currently, academic grants have money allocated for publishing papers. This grant coverage could be expanded to include other types of submissions.

In order to have a sustainable model there would need to be an amount of money paid for each submission, both benchmark and algorithm submission, similar to the paper submission system: once a paper is accepted a certain amount is paid to the publisher to have it published. However, the proposed system would be fairer than that of the journal publisher in that the money earned would be reinvested into platform maintenance and upgrades to support scientists with new features~\citep{mendrikpriv20}. 

\section{How do we train computer scientists to do better science?}
\label{sec:science}

\paragraph{The problem.}

Research communities of all kinds -- from medicine to physics to computer science -- grapple with the mechanics of making sound scientific inferences and educating their students with their community’s norms. Each field has its own approach to rigorous analyses and its own issues with scientific rigor (e.g. reproducibility and replication crises). For example, machine learning, though it conducts experiments as many other fields do, often makes less use of the rigorous statistical methods for drawing scientific conclusions that are more common in fields such as economics, psychology, and medicine. One reason is that as a community we may be failing to train computer scientists to do the kind of research we hope to see at flagship conferences like NeurIPS in the future. Motivated by this issue, a group at the workshop’s brainstorming session asked the question, how should we train computer scientists to do better science?

A significant problem is that many machine learning researchers have never been explicitly educated in research methods or the philosophy of science. For many computer science graduate students, their advisor and fellow graduate students are their exemplars and tutors in how to perform, present, and review ML research, and how their work relates to science as a whole. In practice, advisors may be too busy to explicitly instruct their students in the machinery of science, and students may indirectly and noisily infer from their peers, reviews they receive, and papers they read what the purpose of scientific processes is. That is, they may learn the superficial lesson that papers can only be accepted if they beat the current state-of-the-art, or that it is okay for reviews to be dogmatic, uninformed, and/or conversational. The danger is that there may be little understanding as to why we do research. 

Another aspect complicating educating students about scientific rigor are incentives for faculty advisors, graduate students, and those hoping to get into the field. Machine learning conferences happen frequently throughout the year and there is a pressure to publish multiple papers within the year -- several top machine learning faculty had more than $80$ papers published in 2019. Top universities and companies hire those who have a track record of “first author publications” at top ML conferences (NeurIPS, ICML, etc.). Even before getting into a PhD program in machine learning, there is often an expectation that applicants have published in the past, reinforcing the rush to publish quickly even before enrolling in graduate school. The implicit priority is getting work accepted rather than understanding or advancing science.	As a result, proper scientific practices may be sidelined in the rush to put out more papers -- why learn how to, and then run a suite of sensitivity analyses or ablation studies when such studies may not be necessary for publication given current standards. 

\paragraph{The discussion.}

Discussants looked not only at the problems, but towards possible solutions such as how incentive structures could be changed. For example, ML could learn from other fields and require for promotion or hiring a ``job market paper” that is expected to be very robust and is held to high scientific standards. Indeed, some universities do this already to some extent, focusing on ``best three papers” of candidates.\footnote{See, for example, a recent job posting which requests the ``three papers that best represent their research/experience'': \href{https://www.aclweb.org/portal/content/umass-amherst-computer-science-hiring-faculty-data-science}{https://www.aclweb.org/portal/content/umass-amherst-computer-science-hiring-faculty-data-science}.}

Additionally, reviewer guidelines could be changed to emphasize the robustness of the scientific inferences being made -- though lack of quality reviews certainly contributes to current trends. Another approach would be to switch focus of publications from conferences to journals (as is more common in other fields), where there are more iterations of peer-review and more time to release polished work; however, discussants noted that the perception within ML is that this would slow down the pace of research and may therefore be unlikely to be adopted.

Another concrete suggestion was to create online courses that introduce the scientific method and scientific thinking to machine learning researchers in particular. That is, many newcomers to ML learn through popular online courses (such as through Coursera\footnote{\href{https://www.coursera.org/}{https://www.coursera.org/}} or Stanford’s lectures on Convolutional Neural Networks\footnote{\href{http://cs231n.stanford.edu/}{http://cs231n.stanford.edu/}} that introduce technical aspects of machine learning. However, they do not touch on what makes for good science in ML, which new classes could address. Such classes could additionally discuss current struggles in doing good ML science, such as how the community’s rapid growth has led to low reviewer quality at large conferences. In this way, a course could introduce concepts of what makes for good reviews, and what the whole process of peer review is designed to do. 

The benefit of learning material that is available online is that it is scalable; however, another impactful intervention would be to have such classes (or others in methods or the philosophy of science) as part of the curriculum at universities for CS graduate students. During the discussion, participants remarked that it was strange that they graduated with a Ph.D., a degree designed to prepare them to contribute to the scientific community, without ever having taken a class on the philosophy of science or research methods. Such gaps in the curriculum likely contribute to problematic trends in the ML community.

A final line of discussion related to what ML conferences could do to help contribute to the continuing education of scientists. A discussant noted how useful the ``New in ML workshop'' at NeurIPS\footnote{\href{https://nehzux.github.io/NewInML2019/}{https://nehzux.github.io/NewInML2019/}} could be towards that end. The workshop aimed to help researchers who had not previously published at NeurIPS, and potentially could include talks on methods in research and current problems in the ML landscape. Conferences could include space for workshops (such as the retrospectives workshop) that encourage the community to self-reflect, or to introduce new tracks that encourage papers that provide objective evidence for problematic trends within the community and how to address them.

The conclusion of the session was that right now, as a community we may not be well-preparing the next generation of ML researchers to do the kind of science that we ultimately want reported in an ideal future conference. However, discussants were optimistic that there were practical ways to improve ML education if there was interest and energy to implement them.

\section{Conclusion}
\label{sec:conclusion}

In this report, we summarize ideas and discussion from the NeurIPS 2019 Retrospectives workshop. Our hope is that this report brings awareness to some of the ideas that are currently being discussed for improving the machine learning field as a whole, and contributes towards the goal of better aligning the field with the goal of improving scientific understanding. 

\section*{Acknowledgements}
Thanks to Prof. Yoshua Bengio and Jonathan Frankle. We thank the panelists at the  ML Retrospectives workshop at NeurIPS 2019: Yoshua Bengio, Jonathan Frankle, Melanie Mitchell, Joelle Pineau, Gael Varoquaux (alphabetically ordered by last name), and the organizers of the workshop: Ryan Lowe, Koustuv Sinha, Abhishek Gupta, Jessica Forde, Xavier Bouthillier, Peter Henderson, Michela Paganini, Shagun Sodhani, Kanika Madan, Joel Lehman, Joelle Pineau, and Yoshua Bengio. Additionally, we would like to thank the participants of the brainstorming session at the workshop: Robert Aduviri, Diogo Almeida, Emmanuel Bengio, Wonmin Byeon, Kamil Ciosek, Debajyoti Datta, Jesse Dodge, Ann-Kathrin Dombrowski, Marc Dymetman, Jonathan Frankle, Niklas Gebauer, Barb Hirsch, Rishub Jain, David Jensen, Pan Kessel, Andrey Kurenkov, Shayne Longpre, Kanika Madan, Ilja Manakov, Luke Merrick, Elliot Meyerson, Trishala Neeraj, Andre Pacheco, Michela Paganini, Vipin R. Pillai, Edward Raff, Daniel Seita, and Rupesh Srivastava (alphabetically ordered by last name).


\bibliography{non_style}

\end{document}